\documentclass[reprint, prl, superscriptaddress, amsmath, amssymb,graphicx]{revtex4-2}
\usepackage{braket}
\usepackage{graphicx}
\usepackage{hhline}
\usepackage{multirow}
\usepackage{color}
\usepackage{times}
\usepackage{url}
\usepackage{tikz}
\usepackage{tikz-feynman}
\usepackage{dsfont}
\usepackage{mathtools}
\usepackage{dcolumn}
\usepackage{setspace}
\usepackage{ulem}
\usepackage{physics}
\usepackage{comment}
\usepackage{hyperref}
\usepackage{bbm}
\usepackage[bottom]{footmisc}
\usepackage{siunitx}
\usepackage[toc,page,titletoc]{appendix}
\usepackage{amsfonts}
\usepackage[dvipsnames]{xcolor}

\begin{document}
	\title{Wavefront-Dislocation Evolution via Quadratic Band Touching Annihilation}
    
	
	    \author{Rasoul Ghadimi}
		\email{Ghadimi.rasoul@gmail.com}	
        \thanks{These authors contributed equally.}
        \affiliation{Department of Physics, Hanyang University, Seoul 04763, Korea}
        
        \author{Jaehyeon Ahn}
        \thanks{These authors contributed equally.}
        \affiliation{Department of Physics, Hanyang University, Seoul 04763, Korea}

        \author{Sangmo Cheon}
		\email{sangmocheon@hanyang.ac.kr }
        \affiliation{Department of Physics, Hanyang University, Seoul 04763, Korea}
        \affiliation{High Pressure Research Center, Hanyang University, Seoul 04763, Korea}
        \affiliation{Research Institute for Natural Science, Hanyang University, Seoul 04763, Korea}

	\date{\today}

\begin{abstract}
Wavefront dislocations (WDs)---phase singularities observed in quasiparticle interference (QPI) experiments---have been widely interpreted as the definitive real-space signatures of Berry phases in graphene-family systems. 
Here, we disentangle the roles of topological charge and pseudospin texture in WD experiments.
By investigating various way of the annihilation of quadratic band touchings (QBTs) in bilayer graphene and magneto--spin--orbit graphene systems, we demonstrate that WD evolution is governed exclusively by changes in the underlying pseudospin winding, while remaining insensitive to the topological charge (i.e., vorticity) of the band touching itself.
Our results imply that WD measures wavefunction pseudospin texture rather than a diagnostic of topological charge and provide solid-state platforms in which WD evolution can be engineered and observed.
\end{abstract}

\date{\today}
\maketitle
 
\textit{Introduction.}
Electrons in crystals are fundamentally characterized by their energy dispersion and Bloch wavefunctions~\cite{Solyom2008Fundamentals}.  
While energy dispersions are now routinely mapped by high-resolution angle-resolved photoemission spectroscopy (ARPES)~\cite{ARPESZhang2022}, directly accessing wavefunction information remains substantially more challenging, especially in complex multiband systems.  
This challenge is consequential because nontrivial wavefunctions induce nonzero quantum geometry---including Berry curvature and quantum metric---that governs a wide range of unconventional electronic responses~\cite{RevModPhys.83.407,QGTJiang2025,QGTNagaosa2025,Yu2025}.  
Although wavefunction reconstruction has been demonstrated in specific two-band settings using quantum simulators and optical probes~\cite{PhysRevLett.122.210401,Gianfrate2020,doi:10.1126/science.ado6049,Kang2025,PhysRevResearch.5.L032016}, extending this access to multiband solids remains an outstanding problem.

Scanning tunneling microscopy (STM) in spectroscopy mode (STS) ~\cite{PhysRevB.31.805,PhysRevLett.50.1998} provides a real-space probe of electronic structure~\cite{RevModPhys.59.615,JEON202458}.  
Quasiparticle interference (QPI) in STM--STS experiments exploits impurity-induced LDOS modulations (Friedel oscillations)~\cite{Friedel1952} to access momentum-space scattering information via Fourier transformation (FT--STS)~\cite{doi:10.1126/science.275.5307.1764,Wang2025}.    
Recent studies in graphene-family systems have demonstrated that filtered real-space FT–STS reconstructions can exhibit wavefront dislocations (WDs), where interference wavefronts terminate or merge near an impurity~\cite{Dutreix2019,PhysRevLett.125.116804,Liu2024}.  
In monolayer and bilayer graphene, the observed two and four WDs have been associated with the $\pi$ and $2\pi$ Berry phases of Dirac nodes and quadratic band touchings (QBTs), respectively~\cite{Dutreix2019,PhysRevLett.125.116804}.  
WDs have also been discussed in the context of twisted bilayer graphene (TBG) moir\'e energy bands as a potential diagnostic of the identical vorticity of their moir\'e Dirac nodes~\cite{PhysRevLett.125.176404}. This identical vorticity underlies the fragile topology, or equivalently the topological obstruction to Wannierization, and impacts their correlated electronic instabilities~\cite{PhysRevX.9.021013,PhysRevLett.123.036401}. 
Yet, a fundamental question remains whether WD evolution is governed by the pseudospin texture or by the topological charge itself; disentangling these effects is essential for the correct interpretation of WD experiments.

Here, we address this issue using bilayer binary honeycomb lattices (BBHLs), which provide a tunable platform to annihilate a QBT through two distinct mechanisms.  
By sliding layers from BA to AA configuration, we eliminate the middle-band QBT while significantly reshaping the interlayer pseudospin texture, thereby driving the evolution of WDs.
In contrast, a suitable sublattice potential $m$ removes the same middle-band QBT by transferring it to adjacent bands, but leaves both pseudospin winding and WD patterns unchanged.  
This controlled comparison demonstrates that WDs are determined by the pseudospin texture rather than the topological charge or vorticity of the band touching, even when the same QBT is annihilated. 
The predicted WD phenomenology is directly relevant to sliding bilayer graphene~\cite{Kim2013,26q7-dsm1}, BX bilayers [where B is boron, and X is arsenide (As), phosphide (P), or nitride (N)]~\cite{3pnm-76hh}, metamaterial implementations~\cite{PhysRevB.111.024203}, and magneto--spin--orbit graphene where spin acts as an effective layer degree of freedom~\cite{Marchenko2012,doi:10.1021/nl504693u,doi:10.1021/acs.nanolett.7b01548,PhysRevB.99.085411,PhysRevB.99.195452,PhysRevB.104.155423,sym15020516,ZOU20193162,PhysRevB.108.235166,Zhang2014,doi:10.1021/acs.nanolett.7b01393,PhysRevB.89.075422,Leutenantsmeyer_2017,PhysRevB.74.165310,PhysRevB.82.245412,Farajollahpour2018,3pnm-76hh}.

\begin{figure}[t!]
    \centering
    \includegraphics[width=1\linewidth]{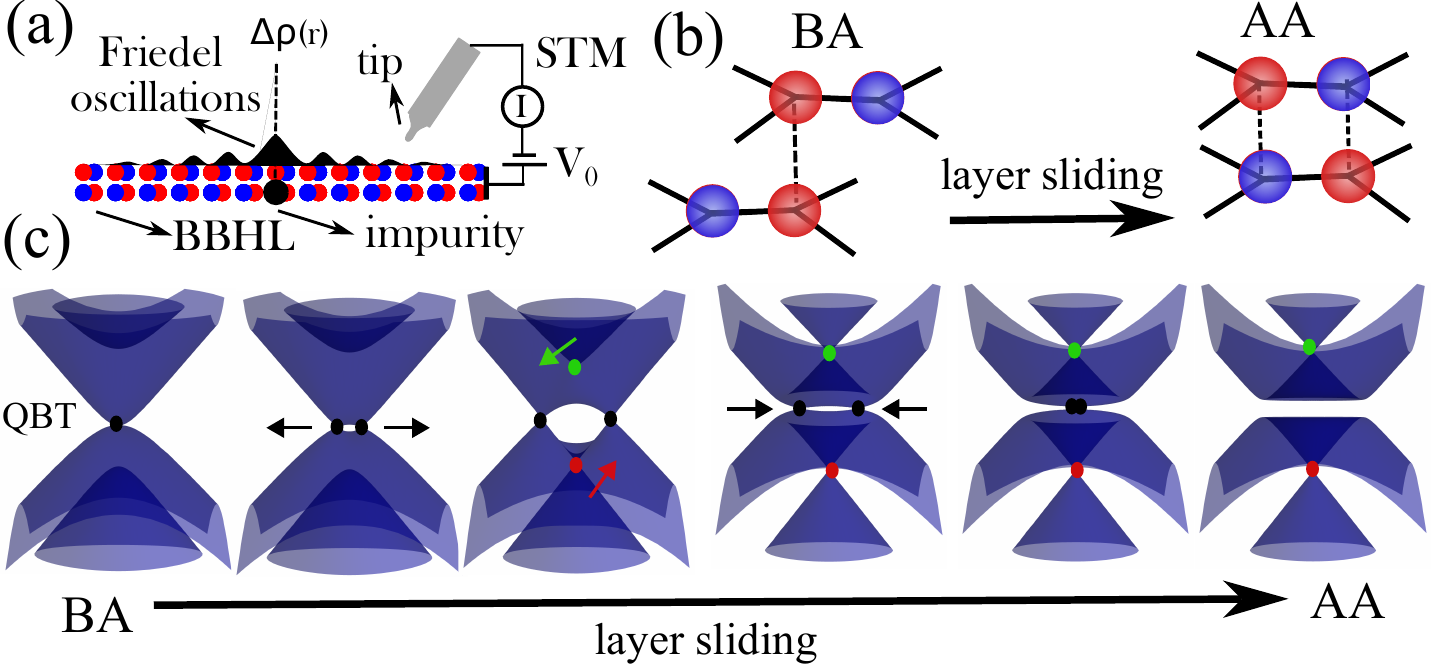}
    \caption{
    \textbf{(a)} STM setup for measuring the impurity-induced LDOS modulation $\Delta\rho(\mathbf r)$ (Friedel oscillations).
    \textbf{(b)} BA and AA stackings of BBHL, connected by layer sliding along a space--time-inversion-symmetric path.
    \textbf{(c)} Evolution of the energy dispersion near the $\mathbf K$ valley under sliding.
    In (c), dots mark touching nodes between the lower (red), middle (black), and upper (green) bands.
    }
    \label{fig:fig1}
\end{figure}

\textit{Model.}
We construct a BBHL model by stacking two honeycomb layers.
Figure~\ref{fig:fig1}(b) shows the unit cell of the two symmetric BA and AA stacking configurations:
AA corresponds to two layers placed directly on top of each other, whereas BA is obtained by sliding one layer along a horizontal nearest-neighbor direction.
The low-energy electronic structure of BBHL with $p_z$ orbitals is effectively dominated by two decoupled valleys, $\mathbf{K}$ and $\mathbf{K}'=-\mathbf{K}$
(see Refs.~\cite{PhysRevB.87.205404,3pnm-76hh} and the \textit{Supplemental Material} (SM)\footnote{See SM at ****.} for the tight-binding model).
The band evolution under sliding near one valley is illustrated in Fig.~\ref{fig:fig1}(c).
The dispersion at the other valley is identical due to inversion $\mathcal{P}=\sigma_x\tau_x \otimes (\mathbf{k}\rightarrow-\mathbf{k})$ and time-reversal symmetries $\mathcal{T}=\mathcal{K} \otimes (\mathbf{k}\rightarrow-\mathbf{k})$,
where $\mathcal{K}$ denotes complex conjugation and $\sigma_x$ ($\tau_x$) exchanges sublattice (layer) indices.

At BA stacking [left panel of Fig.~\ref{fig:fig1}(c)], the energy dispersion exhibits a middle band QBT.  
This QBT carries a vorticity charge (i.e., a patch Euler class  ~\footnotemark[1])   protected by the space-time inversion symmetry $I_{\text{ST}}\equiv\mathcal{PT}$ \cite{PhysRevX.9.021013,Bouhon2020,3pnm-76hh}.
Because $I_{\text{ST}}$ is preserved both under sliding toward AA stacking and under a sublattice potential $m$
[red $+m$ and blue $-m$ in Fig.~\ref{fig:fig1}(b)], an infinitesimal $I_{\text{ST}}$-preserving slide can only split the middle-band QBT into two middle-band Dirac nodes with the same vorticity [black dots in Fig.~\ref{fig:fig1}(c)].
Upon further sliding, additional Dirac nodes in the adjacent bands (green and red dots) approach the vicinity of $\mathbf K$ ($\mathbf K'$).  
These adjacent-band nodes flip the relative vorticity of the middle-band Dirac nodes, via non-Abelian charge conservation~\cite{3pnm-76hh}.  
By further sliding with a small $m\neq0$, the middle-band Dirac nodes meet and annihilate due to their opposite relative vorticity.  
Consequently, at AA stacking [right panel of Fig.~\ref{fig:fig1}(c)], the middle band is gapped while Dirac nodes appear in the adjacent bands.  
Alternatively, in the BA configuration, a sufficient sublattice potential $m$ removes the middle-band QBT by transferring it to either the lower or upper band, depending on the sign of $m$ [see Fig.~\ref{fig:fig4}(a)] ~\cite{3pnm-76hh}.  
These two symmetry-preserving routes remove the same QBT while affecting the pseudospin texture differently, providing a controlled setting to identify what WD actually measures.

\begin{table}[b]
    \caption{
    WD charges $n_{l'\sigma',l\sigma}$ at BA and AA stackings for STM $l'\sigma'$ and impurity $l\sigma$ (IMP)  channels.
    Left (right) block lists intralayer (interlayer) channels, showing that only interlayer channels exhibit WD charge evolution between BA and AA.
    }
\label{tab:wavefrontdislocation}
\centering
\begin{tabular}{cc|ccc@{\hspace{1.5cm}}cc|cc}
STM & IMP & BA & AA && STM & IMP & BA & AA \\
\cline{1-4}\cline{6-9}
2A & 2A &  0 &  0 && 2A & 1A &  2 &  0 \\
2B & 2A &  2 &  2 && 2B & 1A &  4 &  2 \\
2A & 2B & -2 & -2 && 2A & 1B &  0 & -2 \\
2B & 2B &  0 &  0 && 2B & 1B &  2 &  0 \\
\end{tabular}
\end{table}

\textit{Wavefront Dislocation.}  
Impurity-induced Friedel oscillations and their Fourier-transformed signatures in QPI experiments have been extensively studied in graphene-family materials and beyond~\cite{Crommie1993,PhysRevB.107.L041404,PhysRevB.82.193405,PhysRevB.88.205416,PhysRevB.81.045409,PhysRevLett.120.106801,PhysRevLett.100.076601,PhysRevB.93.035413,PhysRevB.86.045444,PhysRevLett.101.206802,10.1063/1.4890543,PhysRevB.87.245413,PhysRevB.111.134505,PhysRevB.107.235423}.  
In our bilayer setting, the impurity-induced LDOS modulation $\rho_{l'\sigma',l\sigma}(\omega,\mathbf{r})$ depends on the energy $\omega$, the impurity site $l\sigma$, the STM tip position $l'\sigma'$, and their separation $\mathbf{r}$. 
Here $\sigma,\sigma' = A,B$ denote sublattices and $l,l' = 1,2$ label the bottom and top layers, respectively. Experimentally, STM typically probes only the top layer ($l'=2$). 
In the following, we call each pair of STM and impurity sublattice indices $(l'\sigma',l\sigma)$ as a scattering channel.

\begin{figure}[]
    \centering
    \includegraphics[width=0.85\linewidth]{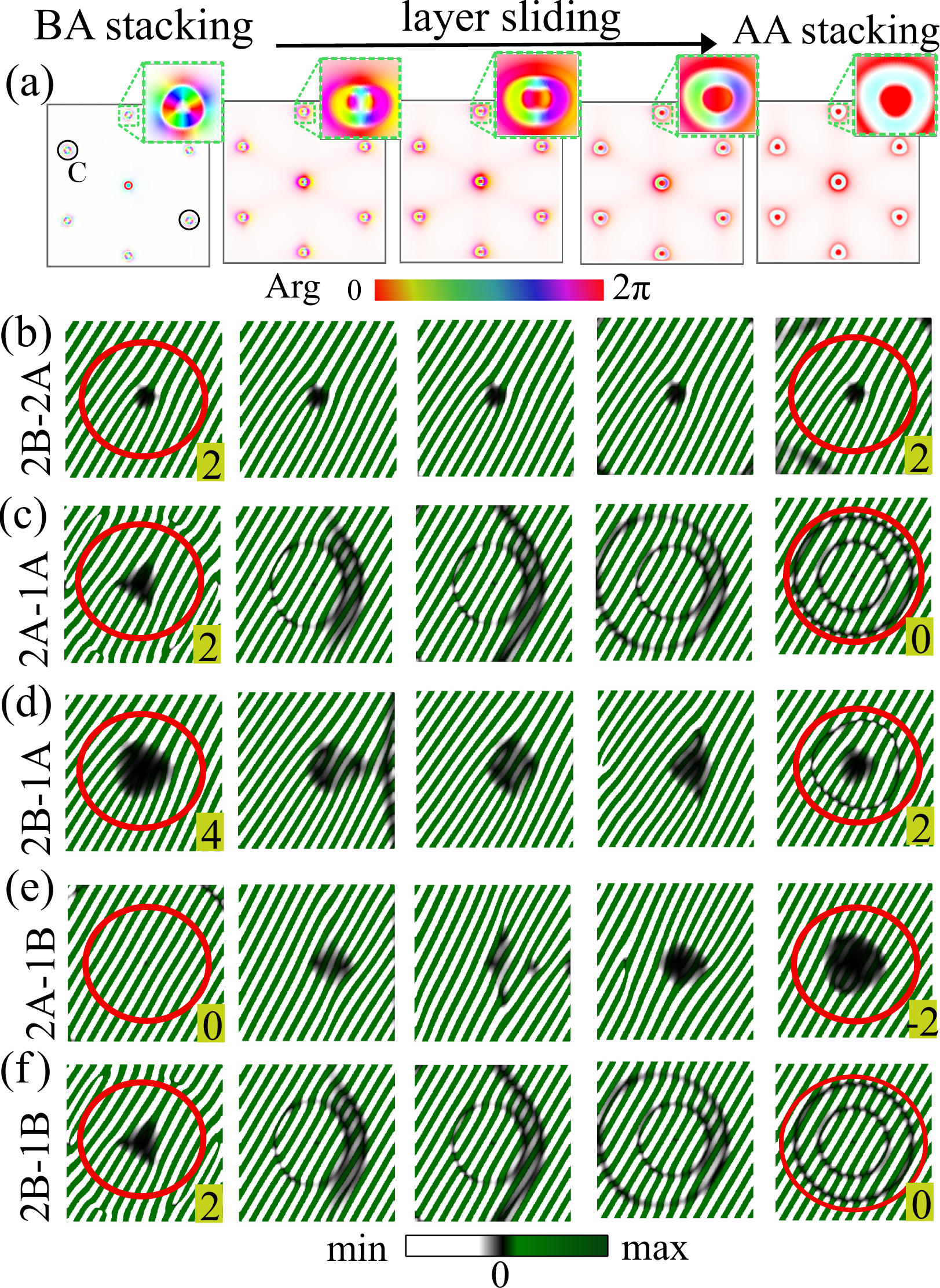}
    \caption{
    FT--STS analysis of QBT annihilation during the layer sliding. (a) FT--STS map $\rho_{2A,1A}(\omega,\mathbf q)$ during the sliding.   The color indicate phase; insets show the zoomed image. ``C'' marks the chosen pair used for filtering. (b--f) Corresponding intervalley-filtered real-space modulations $\rho^{\text{filtered}}_{l'\sigma',l\sigma}(\omega,\mathbf r)$ for representative stacking configurations; left labels indicate the STM  and impurity  channel. 
    Numbers in the green boxes denote the WD charge, obtained by taking the difference between the numbers of green wavefront lines coming or going from the red circle.
    For plotting we use $\omega=0.2~\mathrm{eV}$ above half filling and $m=0.05~\mathrm{eV}$.
    }
    \label{fig:fig2}
\end{figure}

Taking the Fourier transformation $\rho_{l'\sigma',l\sigma}(\omega, \mathbf{q})$ uncovers momentum-space scattering characteristics that are otherwise masked in the raw real-space STS data. 
As an example, Fig.~\ref{fig:fig2}(a) shows the numerically obtained FT--STS map $\rho_{2A,1A}(\omega, \mathbf{q})$ along the sliding path. 
The map exhibits a dominant central peak corresponding to intravalley scattering, surrounded by six distinct satellite features originating from intervalley scattering processes~\cite{Dutreix2019,PhysRevLett.125.116804}. 
To extract the wavefunction information encoded in a specific scattering channel, we apply a window in $\mathbf{q}$-space 
and perform a filtered inverse Fourier transformation. 
This procedure yields the filtered real-space modulation $\rho^{\text{filtered}}_{l'\sigma',l\sigma}(\omega, \mathbf{r})$, which isolates the spatial interference pattern associated with the chosen intervalley process.
Specifically, we choose a circular window centered at $\pm\Delta\mathbf{K}$ [region ``C'' in Fig.~\ref{fig:fig2}(a)], for which $\rho^{\text{filtered}}_{l'\sigma',l\sigma}(\omega,\mathbf{r})$ exhibits wavefronts perpendicular to $\Delta\mathbf{K}$ [Fig.~\ref{fig:fig2}(b--f)].  
WDs then correspond to the net termination and merging of these wavefronts near the impurity.  
In the following, we track the evolution of these WDs under sliding and sublattice tuning to determine whether WD is fundamentally controlled by vorticity or by pseudospin texture.

\textit{WD under Sliding.}
Since sliding primarily reshapes interlayer hybridization, it reorganizes the multicomponent wavefunction (or pseudospin) texture, and this reorganization should be encoded in the WD patterns extracted from intervalley-filtered QPI data.

For the high-symmetry BA and AA stackings, our analytic derivation reveals that the filtered modulation $\rho^{\text{filtered}}_{l'\sigma',l\sigma}(\omega,\mathbf r)$ takes the approximate form
\begin{align}
f(\omega,r)\cos\!\left(\Delta\mathbf{K}\!\cdot\!\mathbf{r}
+ n_{l'\sigma',l\sigma}\theta_r
+ \phi_{l'\sigma',l\sigma}\right),
\label{equ:gen}
\end{align}
where $\mathbf{r}=r(\cos\theta_r,\sin\theta_r)$, $\phi_{l'\sigma',l\sigma}$ is a channel-dependent constant phase, and $f(\omega,r)$ is a radial envelope that depends on $(m,\omega,r)$ and stacking but is independent of $\theta_r$.
The integer $n_{l'\sigma',l\sigma}$ defines the WD charge: encircling the impurity ($\theta_r:0\!\to\!2\pi$) yields a phase winding $2\pi n_{l'\sigma',l\sigma}$, read out as a net dislocation \footnotemark[1].

Table~\ref{tab:wavefrontdislocation} lists the WD charges $n_{l'\sigma',l\sigma}$ for BA and AA stackings.
A key consequence is that layer sliding modifies $n_{l'\sigma',l\sigma}$ exclusively for interlayer channels ($l\neq l'$), while intralayer channels ($l=l'$) remain unchanged, consistent with the fact that sliding predominantly affects interlayer coupling.
Away from the high-symmetry stackings, Eq.~(\ref{equ:gen}) no longer holds in a simple closed form because the envelope $f(\omega,r)$ acquires an explicit $\theta_r$ dependency.
We therefore track WD evolution numerically in Fig.~\ref{fig:fig2}(b--f), identifying the dislocation by counting the difference of wavefront lines coming or going from the red circle surrounding the impurity (numbers in each panel).
The numerical results confirm the results in Table~\ref{tab:wavefrontdislocation}: WD evolves under sliding for only interlayer channels ($l\neq l'$) [cf. Fig.~\ref{fig:fig2}(b) with Fig.~\ref{fig:fig2}(c--f)].
Therefore, BBHLs provide a solid-state platform to engineer and track WD evolution, in close analogy with classical water-wave interference experiment~\cite{MVBerry_1980}.
Notably, this sliding-induced WD evolution is nontrivial: WDs were reported to be largely insensitive to trigonal warping and to both topological and non-topological gaps~\cite{PhysRevB.103.L161407,wavefrontDislocationTrigonalwrapping,PhysRevB.104.035402}, whereas sliding reorganizes multicomponent wavefunction texture and thereby reshapes the WD pattern.

\begin{figure}[t!]
    \centering
    \includegraphics[width=0.85\linewidth]{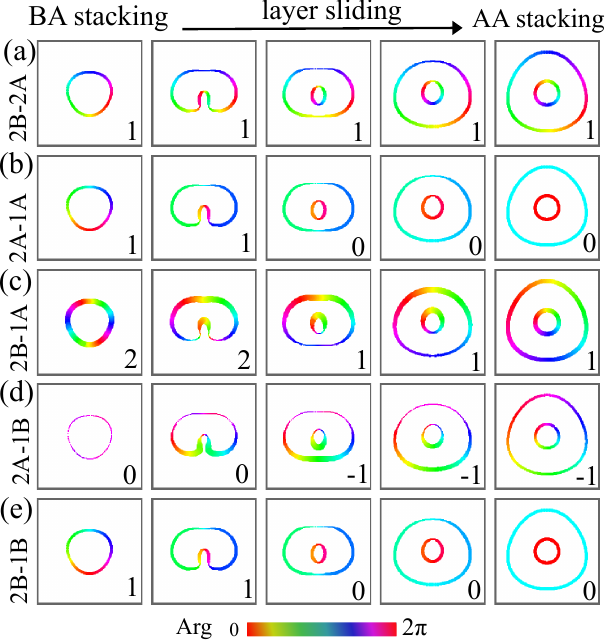}
    \caption{
    Pseudospin textures versus sliding.
    (a--e)
    Evolution of $\mathbf{s}_{l'\sigma',l\sigma}$ along the constant-energy contour at $\omega=0.2~\mathrm{eV}$ for different stackings.
    The magnitude and phase of $\mathbf{s}_{l'\sigma',l\sigma}$ are encoded by the line thickness and color, respectively.
    Left labels specify STM and impurity channel, and numbers denote the pseudospin winding, which is unchanged for $l'=l$ but evolves for $l'\neq l$.
    }
    \label{fig:fig3}
\end{figure}

\textit{Pseudospin Winding.}   
While we have demonstrated the evolution of WDs during layer sliding in BBHL, interpreting these results solely through the lens of energy dispersion is insufficient. This is because different scattering channels $\rho^{\text{filtered}}_{l'\sigma',l\sigma}$ can exhibit distinct behaviors. Following previous studies, pseudospin \cite{PhysRevB.84.205440} provides a natural framework for interpreting WDs in graphene-family systems~\cite{PhysRevB.86.045444}. For the BBHL, this notion must be generalized to account for the four sublattice--layer degrees of freedom.

We construct a generalized pseudospin by selecting the corresponding components of the Bloch wavefunction $\ket{\psi_0(\mathbf{k})}$ for each channel $(l'\sigma',l\sigma)$ [see more details in SM].
In other words,
$
\ket{\psi(\mathbf{k})}_{l'\sigma',l\sigma}
\equiv
\big(\langle l'\sigma'|\psi_0(\mathbf{k})\rangle,\ \langle l\sigma|\psi_0(\mathbf{k})\rangle\big),
$
where $\ket{l,\sigma}$ denotes the state on layer $l$ and sublattice $\sigma$.
Let $s_x$ and $s_y$ be the Pauli matrices acting on the $(l'\sigma',l\sigma)$ subspace.  
We then define the generalized in-plane pseudospin texture as
\begin{equation}
    \mathbf{s}_{l'\sigma',l\sigma}
    \equiv \big(\langle s_x \rangle_{l'\sigma',l\sigma}, \langle s_y \rangle_{l'\sigma',l\sigma} \big),
\end{equation}
with expectation values taken with respect to $\ket{\psi(\mathbf{k})}_{l'\sigma',l\sigma}$.
The direction of $\mathbf{s}_{l'\sigma',l\sigma}$ gives the pseudospin angle, while its magnitude reflects the relative weight of the selected components in $\ket{\psi_0(\mathbf{k})}$.

Figure~\ref{fig:fig3} illustrates the evolution of the pseudospin texture under sliding on a constant-energy contour.
We find that the pseudospin winding depends on both the scattering channel (STM and impurity) and the stacking configuration. 
Interestingly, the pseudospin winding remains unchanged for intralayer channels~[Fig.~\ref{fig:fig3}(a)], whereas it evolves under sliding for interlayer channels~[Figs.~\ref{fig:fig3}(b--e)].  
The change of the pseudospin winding occurs through merging and splitting of the constant-energy contours.

In the SM~\footnotemark[1], our analysis demonstrates that WD is related to pseudospin winding.
For sliding BBHL with time-reversal symmetry $\mathcal{T}=\mathcal{K}\otimes (\mathbf k\rightarrow-\mathbf k)$, the pseudospin winding in the two valleys is always opposite, so WD charge equals twice the pseudospin winding.
For instance, the two WDs in $\rho^{\text{filtered}}_{2B,2A}$ arise from a single winding of $\mathbf{s}_{2B,2A}$ on the energy contour, and both remain unchanged with sliding [cf.\ Figs.~\ref{fig:fig2}(b) and \ref{fig:fig3}(a)]. 
In contrast, the four WDs in $\rho^{\text{filtered}}_{2B,1A}$ originate from a double winding of $\mathbf{s}_{2B,1A}$ in BA stacking; both the WD number and the winding are reduced by half in AA stacking [Figs.~\ref{fig:fig2}(d) and \ref{fig:fig3}(c)].

\textit{Sublattice Tuning.}
As a controlled alternative to sliding, we investigate removing the middle-band QBT in the BA configuration by introducing a sublattice potential $m$.
Experimentally, such a modulation of $m$ can be realized by varying the binary-atom composition, e.g., in BX bilayers~\cite{3pnm-76hh}.
Figure~\ref{fig:fig4}(a) shows that tuning $m$ transfers the QBT from the middle band at $m=0$ to the lower band for $m=-0.3~\mathrm{eV}$ and to the upper band for $m=0.3~\mathrm{eV}$.
In Figs.~\ref{fig:fig4}(b) and \ref{fig:fig4}(c), we plot $\mathbf{s}_{2A,1A}$ on constant-energy contours at $\omega=-0.5~\mathrm{eV}$ and $\omega=0.5~\mathrm{eV}$, respectively.
We see that the winding of $\mathbf{s}_{2A,1A}$ remains invariant for both $\omega=\pm0.5$ eV and different $m$, notwithstanding the energy-dispersion evolution.
Consistently, in Figs.~\ref{fig:fig4}(d--e), we plot $\rho^{\text{filtered}}_{2A,1A}$ for different $\omega$ and $m$ and observe two WDs that are insensitive to both parameters.
This behavior is also captured by Eq.~(\ref{equ:gen}), using the same $n_{l'\sigma',l\sigma}$ in Table~\ref{tab:wavefrontdislocation}: varying $m$ primarily renormalizes the radial envelope $f(\omega,r)$ and therefore leaves the WD charge unchanged.
Thus, transferring the QBT between bands via $m$ leaves WD unchanged, providing a controlled null test that WD tracks pseudospin winding rather than band-touching vorticity.

\textit{Role of Vorticity.}  
In the BBHL, removing the middle-band QBT---either via layer sliding or a sufficient sublattice potential---involves a non-Abelian charge conversion that trivializes the QBT vorticity, as discussed in Ref.~\cite{3pnm-76hh}.  
Yet, as shown in Fig.~\ref{fig:fig4}, the filtered patterns at $\omega=-0.5~\mathrm{eV}$ for $m=0.3~\mathrm{eV}$ and at $\omega=0.5~\mathrm{eV}$ for $m=-0.3~\mathrm{eV}$ exhibit the same WDs even though these energies do not intersect the band hosting the transferred QBT.  
This controlled comparison shows that band-touching vorticity and the associated topological charges, similar to strong topological indices \cite{PhysRevB.104.035402}, do not determine the WD pattern.  
Instead, comparing Figs.~\ref{fig:fig2}--\ref{fig:fig4}, we conclude that WDs are set by the underlying pseudospin textures.

\begin{figure}[t!]
    \centering
    \includegraphics[width=0.85\linewidth]{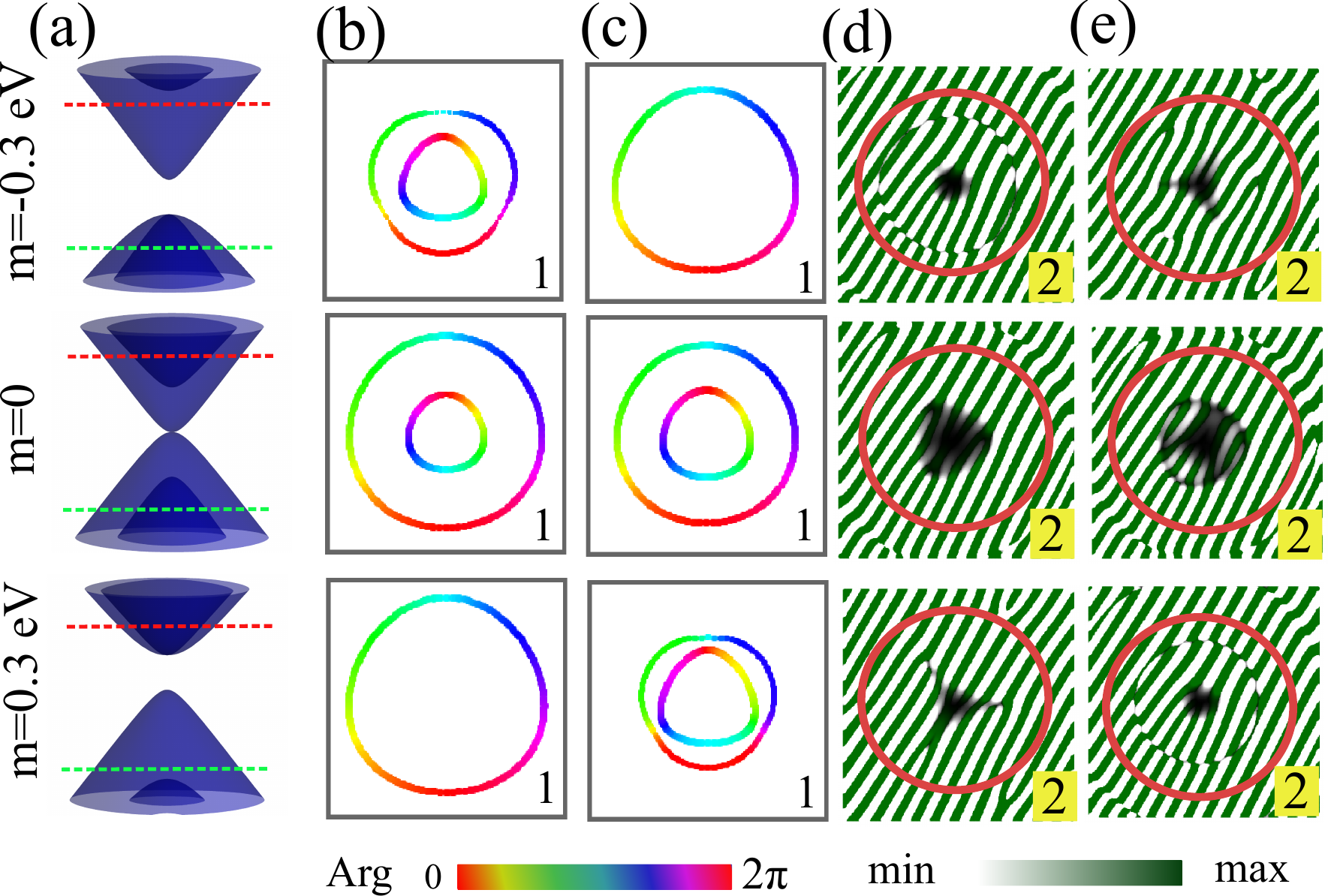}
    \caption{
    WD patterns and pseudospin textures for QBT removal by a sublattice potential $m$.
    (a) Energy dispersion for different $m$; dashed lines mark $\omega=\pm 0.5~\mathrm{eV}$.
    (b,c) $\mathbf{s}_{2A,1A}$ on the corresponding constant-energy contours at (b) $\omega=-0.5~\mathrm{eV}$ and (c) $\omega=0.5~\mathrm{eV}$, and (d,e) $\rho^{\text{filtered}}_{2A,1A}$ at the same energies.
    Numbers in (b,c) indicate the pseudospin winding, while those in (d,e) denote the corresponding WD charge. 
    Transferring the QBT to adjacent bands leaves both pseudospin windings and WDs unchanged.
    }
    \label{fig:fig4}
\end{figure}

\textit{Discussion and Generalization.}
Our results establish that WD evolution is not determined by topological charges.
Consequently, maintaining a strict $I_{\text{ST}}$-symmetric stacking configuration is not essential.
Inducing a small gap at band touchings by breaking $I_{\text{ST}}$ symmetry does not qualitatively alter the pseudospin winding and WD. 
This relaxes material and symmetry constraints and broadens the range of realizable platforms for observing WD evolution.
Experimentally, layer sliding has already been demonstrated in bilayer graphene, for example, via small twisting or bending across a nanoridge~\cite{26q7-dsm1,Kim2013}.  
Large-angle commensurate twisted bilayer graphene provides another promising arena~\cite{PhysRevB.108.L121405}. 
It can exhibit dispersions reminiscent of sliding BBHL, and offering experimentally accessible sliding degrees of freedom~\cite{PhysRevB.108.L121405,PhysRevB.81.161405,PhysRevLett.123.216803,PhysRevLett.133.196603}.  
Furthermore, honeycomb lattices and BBHL-type models have been implemented in metamaterial platforms~\cite{PhysRevB.111.024203,PhysRevB.103.064304,Gardezi_2021,PhysRevB.103.214311,PhysRevB.102.180304,PhysRevApplied.17.034061}, where Friedel-like oscillations can be used to observe WD evolution driven by tunable couplings~\cite{Dutreix2021}.  

Spin--orbit coupling (SOC) provides additional knobs in graphene-based platforms. 
Because SOC is typically weak in pristine graphene, the two spin species contribute identically to $\rho^{\text{filtered}}$.
Consequently, for a non-magnetic impurity in the absence of magnetic interactions, the WD patterns in spinful BBHL models remain invariant and are identical to those of the spinless case.

On the other hand, SOC can be proximity-induced and may qualitatively modify WD experiments.  
Specifically, BBHL-type physics can be realized in proximitized magneto--spin--orbit-coupled graphene~\cite{Marchenko2012,doi:10.1021/nl504693u,doi:10.1021/acs.nanolett.7b01548,PhysRevB.99.085411,PhysRevB.99.195452,PhysRevB.104.155423,sym15020516,ZOU20193162,PhysRevB.108.235166,Zhang2014,doi:10.1021/acs.nanolett.7b01393,PhysRevB.89.075422,Leutenantsmeyer_2017,PhysRevB.74.165310,PhysRevB.82.245412,Farajollahpour2018},  
where spin effectively plays the role of the layer degree of freedom in the BBHL model (see Ref.~\cite{3pnm-76hh} and the SM).  
The energy dispersion of Rashba-coupled monolayer graphene hosts QBTs analogous to BA-stacked bilayers, whereas a purely in-plane exchange field yields a dispersion analogous to AA stacking. 
Therefore, tuning proximity conditions to enhance the exchange field relative to Rashba SOC provides a mechanism to annihilate the QBT.
In this system, WDs can be accessed via spin-polarized STM through magnetic Friedel oscillations induced by a magnetic impurity~\cite{RevModPhys.81.1495,MBode_2003,PhysRevLett.133.036204}.  
Unlike the time-reversal-symmetric sliding BBHL, magnetic interactions generally break $\mathcal{T}$, so each valley spin/pseudospin winding need not be opposite, and WDs are determined by both valleys' pseudospin windings. 
Interestingly, we show that, in contrast to sliding BBHL, purely Rashba-coupled and purely in-plane Zeeman-coupled graphene can nevertheless exhibit the same WD patterns due to their distinct valley spin textures [see SM].
This further underscores that WDs primarily probe spin/pseudospin texture rather than the underlying topological charges.

\section*{Acknowledgments}
We thank Prof. Moon Jip Park and Prof. Tae-Hwan Kim for the discussion.
This work was supported by the National Research Foundation of Korea (NRF) funded by the Ministry of Science and ICT (MSIT), South Korea (Grants No. NRF-2022R1A2C1011646,  NRF-2022M3H3A1085772, RS-2024-00416036, and RS-2025-03392969).
This work was supported by the Creation of the Quantum Information Science R\&D Ecosystem (Grant No. RS-2023-NR068116) through the National Research Foundation of Korea (NRF) funded by the  Korean government (Ministry of Science and ICT).
This work was also supported by the Quantum Simulator Development Project for Materials Innovation through the NRF funded by the MSIT, South Korea (Grant No. RS-2023-NR119931).

\begin{acknowledgements}
\end{acknowledgements}
 \bibliography{ref}

\end{document}